\documentclass{aastex62}
\begin{document}

\title{The 2175 \AA\ extinction feature in the optical afterglow spectrum of GRB\,180325A at $z=2.25$\thanks{Based on observations made with the Nordic Optical Telescope, operated by the Nordic Optical Telescope Scientific Association at the Observatorio del Roque de los Muchachos, La Palma, Spain, of the Instituto de Astrofisica de Canarias. Based on observations collected at the European Organisation for Astronomical Research in the Southern Hemisphere under ESO programme 0100.D$-$0649(A).}
}

\correspondingauthor{T. Zafar}
\email{tayyaba.zafar@aao.gov.au}
\author{T. Zafar}
\affiliation{Australian Astronomical Observatory, PO Box 915, North Ryde, NSW 1670, Australia; tayyaba.zafar@aao.gov.au}

\author{K.~E. Heintz}
\affiliation{Centre for Astrophysics and Cosmology, Science Institute, University of Iceland, Dunhagi 5, 107 Reykjav\'ik, Iceland}
\affiliation{The Cosmic Dawn Center, Niels Bohr Institute, University of Copenhagen, Juliane Maries Vej 30, DK-2100 Copenhagen \O, Denmark}

\author{J.~P.~U. Fynbo}
\affiliation{The Cosmic Dawn Center, Niels Bohr Institute, University of Copenhagen, Juliane Maries Vej 30, DK-2100 Copenhagen \O, Denmark}

\author{D. Malesani}
\affiliation{Dark Cosmology Centre, Niels Bohr Institute, University of Copenhagen, Juliane Maries Vej 30, DK-2100 Copenhagen \O, Denmark}

\author{J. Bolmer}
\affiliation{European Southern Observatory, Alonso de C\'ordova 3107, Vitacura, Casilla 19001, Santiago de Chile, Chile}

\author{C. Ledoux}
\affiliation{European Southern Observatory, Alonso de C\'ordova 3107, Vitacura, Casilla 19001, Santiago de Chile, Chile}

\author{M. Arabsalmani}
\affiliation{Universit\'e Paris Diderot, AIM, Sorbonne Paris Cit\'e, CEA, CNRS, F-91191 Gif-sur-Yvette, France}

\author{L. Kaper}
\affiliation{Astronomical Institute Anton Pannekoek, University of Amsterdam, Science Park 904, 1098 XH, Amsterdam, the Netherlands.}

\author{S. Campana}
\affiliation{INAF - Osservatorio astronomico di Brera, Via E. Bianchi 46, Merate (LC) I-23807, Italy}

\author{R.~L.~C. Starling}
\affiliation{Department of Physics and Astronomy, University of Leicester, University Road, Leicester, LE1 7RH, UK}

\author{J. Selsing}
\affiliation{Dark Cosmology Centre, Niels Bohr Institute, University of Copenhagen, Juliane Maries Vej 30, DK-2100 Copenhagen \O, Denmark}

\author{D.~A. Kann}
\affiliation{Instituto de Astrof\'isica de Andaluc\'ia (IAA-CSIC), Glorieta de la Astronom\'ia, s/n, E18008 Granada, Spain}

\author{A. de Ugarte Postigo}
\affiliation{Dark Cosmology Centre, Niels Bohr Institute, University of Copenhagen, Juliane Maries Vej 30, DK-2100 Copenhagen \O, Denmark}
\affiliation{Instituto de Astrof\'isica de Andaluc\'ia (IAA-CSIC), Glorieta de la Astronom\'ia, s/n, E18008 Granada, Spain}

\author{T. Schweyer}
\affiliation{Max-Planck Institut f$\ddot{u}$r Extraterrestrische Physik, Giessenbachstr. 1, 85748 Garching, Germany}
\affiliation{Technische Universit$\ddot{a}$t M$\ddot{u}$nchen, Physik Dept., James-Franck-Str., 85748 Garching, Germany}

\author{L. Christensen}
\affiliation{Dark Cosmology Centre, Niels Bohr Institute, University of Copenhagen, Juliane Maries Vej 30, DK-2100 Copenhagen \O, Denmark}

\author{P. M\o ller}
\affiliation{European Southern Observatory, Karl-Schwarzschild-Strasse 2, 85748, Garching, Germany}

\author{J. Japelj}
\affiliation{Astronomical Institute Anton Pannekoek, University of Amsterdam, Science Park 904, 1098 XH, Amsterdam, the Netherlands.}

\author{D. Perley}
\affiliation{Astrophysics Research Institute, Liverpool John Moores University, 146 Brownlow Hill, IC2 Liverpool Science Park, Liverpool, L3 5RF, UK}

\author{N.~R. Tanvir}
\affiliation{Department of Physics and Astronomy, University of Leicester, University Road, Leicester, LE1 7RH, UK}

\author{P. D'Avanzo}
\affiliation{INAF-Osservatorio Astronomico di Brera, Via Bianchi 46, I-23807, Merate (LC), Italy}

\author{D.~H. Hartmann}
\affiliation{Department of Physics and Astronomy, Clemson University, Clemson, SV 29634-0978, USA}

\author{J. Hjorth}
\affiliation{Dark Cosmology Centre, Niels Bohr Institute, University of Copenhagen, Juliane Maries Vej 30, DK-2100 Copenhagen \O, Denmark}

\author{S. Covino}
\affiliation{INAF / Brera Astronomical Observatory, Via Bianchi 46, 23807, Merate (LC), Italy}

\author{B. Sbarufatti}
\affiliation{Department of Astronomy and Astrophysics, The Pennsylvania State University, 525 Davey Lab, University Park, PA 16802, USA}

\author{P. Jakobsson}
\affiliation{Centre for Astrophysics and Cosmology, Science Institute, University of Iceland, Dunhagi 5, 107 Reykjav\'ik, Iceland}

\author{L. Izzo}
\affiliation{Instituto de Astrof\'isica de Andaluc\'ia (IAA-CSIC), Glorieta de la Astronom\'ia, s/n, E18008 Granada, Spain}

\author{R. Salvaterra}
\affiliation{INAF - IASF/Milano, via Bassini 15, I-20133 Milano, Italy}

\author{V. D'Elia}
\affiliation{Space Science Data Center - Agenzia Spaziale Italiana, via del Politecnico, s.n.c., I-00133, Roma, Italy}
\affiliation{INAF - Astronomical Observatory of Rome, via Frascati 33, Monte Porzio Catone I-00040, Roma, Italy}

\author{D. Xu}
\affiliation{National Astronomical Observatories, Chinese Academy of Sciences, Beijing 100012, China}

\begin{abstract}
The UV extinction feature at 2175\,\AA\ is ubiquitously observed in the Galaxy but is rarely detected at high redshifts. Here we report the spectroscopic detection of the 2175\,\AA\ bump on the sightline to the $\gamma$-ray burst (GRB) afterglow GRB 180325A at $z=2.2486$, the only unambiguous detection over the past ten years of GRB follow-up, at four different epochs with the Nordic Optical Telescope (NOT) and the Very Large Telescope (VLT)/X-shooter. Additional photometric observations of the afterglow are obtained with the Gamma-Ray burst Optical and Near-Infrared Detector (GROND). We construct the near-infrared to X-ray spectral energy distributions (SEDs) at four spectroscopic epochs. The SEDs are well-described by a single power-law and an extinction law with $R_V\approx4.4$, $A_V\approx1.5$, and the 2175\,\AA\ extinction feature. The bump strength and extinction curve are shallower than the average Galactic extinction curve. We determine a metallicity of [Zn/H]$>-0.98$ from the VLT/X-shooter spectrum. We detect strong neutral carbon associated with the GRB with equivalent width of $W_{\mathrm{r}}(\lambda 1656) =0.85\pm0.05$. We also detect optical emission lines from the host galaxy. Based on the H$\alpha$ emission line flux, the derived dust-corrected star-formation rate is $\sim46\pm4$\,M$_\odot$yr$^{-1}$ and the predicted stellar mass is log\,M$_\ast$/M$_\odot\sim9.3\pm0.4$, suggesting the host galaxy is amongst the main-sequence star-forming galaxies. 
\end{abstract}

%
\section{Introduction\label{introduction}}
Dust plays an important role in the formation and evolution of galaxies. Extinction curves as a function of wavelength provide information on dust grain sizes and  compositions \citep{draine03}. Long-duration $\gamma$-ray bursts (GRBs) are powerful probes to study absolute extinction curves of distant star-forming galaxies because of their bright afterglow emission, simple spectral shape \citep{sari98}, association with short-lived massive stars \citep[e.g.,][]{cano17}, and location in star-forming regions \citep[e.g.,][]{fynbo00,starling11}. Broadband spectroscopy of GRB afterglows allows us to study individual line-of-sight absolute extinction curves \citep{zafar11,zafar18}, and provides  information about chemical abundances \citep{cucchiara15} and gas kinematics \citep{arabsalmani15} in their host galaxies. 

A characteristic feature in the Milky Way (MW) extinction curve is an absorption bump centered at 2175\,\AA, first discovered by \citet{stecher65}. The feature is ubiquitously seen in MW extinction curves \citep{fm07}. The bump strength drops in the interstellar extinction curves of the Large Magellanic Cloud (LMC; \citealt{gordon03}) and is only seen in a few sightlines towards the Small Magellanic Cloud (SMC; \citealt{gordon03,apellaniz12}). The exact origin of the feature is still unclear, although some candidates, such as non-graphitic carbon \citep{mathis94} or polycyclic aromatic hydrocarbons (PAHs; \citealt{draine03}) have been proposed. The reduction in extinction bump strength for different environments is attributed to low metallicities \citep{fitzpatrick04} or different radiative environments \citep{mattsson08}.

Beyond the local Universe, the 2175\,\AA\ extinction feature becomes very uncommon \citep[e.g.,][]{zeimann15,reddy18}. At $z>1$, the extinction bump has been detected in the Great Observatories Origins Deep Survey (GOODS)-\emph{Herschel} field galaxies \citep{buat11}. A sample of massive star-forming galaxies at $z\sim$2 indicates 2175\,\AA\ bumps in their spectra \citep{noll07}. Weaker bumps for active galaxies have been found from the SEDs of $0.5<z<2.0$ galaxies \citep{kriek13}. In a sample of $2 \le z \le 6.5$ COSMOS galaxies, a 2175\,\AA\ bump is reported based on photometry \citep{scoville15}. The extinction feature is also detected in several metal-line absorbing systems towards quasar sightlines \citep[e.g.,][]{jiang11,ma18}. In GRBs, a significant spectroscopic 2175\,\AA\ bump has been confirmed only in four GRBs (GRB\,070802: \citealt{kruhler08,eliasdottir09}, GRB\,080607: \citealt{prochaska09,perley11}, GRB\,080605 and GRB\,080805: \citealt{zafar12}). The peculiar extinction curve towards GRB\,140506A does not seem to be caused by an extreme version of the 2175\,\AA\ extinction feature, but a different phenomenon, the nature of which remains unknown \citep{fynbo14,heintz17}. In general, by studying a $\gamma$-ray flux limited sample, \citet{covino13} found that the 2175\,\AA\ feature is required in at least $15$\% of cases to account for the broadband photometric afterglow SED.

Here, we report another significant {\it spectroscopic} detection of a 2175\,\AA\ bump, discovered in the optical spectrum of GRB\,180325A, nearly a decade since the last reported unambiguous examples. This dearth is despite the powerful current generation of instruments, in particular, the Very Large Telescope (VLT)/X-shooter \citep{vernet11}. The VLT/X-shooter enables near-infrared (NIR) to the ultraviolet (UV) observations in one-shot and is ideally suited to discover such a feature. The Letter is organized as follows: In \S\ref{observations} we describe multi-wavelength observations of the GRB\,180325A afterglow carried out with different instruments. In \S\ref{results} we present our spectral energy distribution (SED) and optical spectrum analysis results. A discussion is provided in \S\ref{discussion} and conclusions in \S\ref{conclusions}.

%
%
\section{Observations and SED construction}\label{observations} 
At 01:53:02\,UT on 2018 March 25, the \emph{Neil Gehrels Swift Observatory} detected GRB\,180325A with its Burst Alert Telescope (BAT) and its X-Ray Telescope (XRT) began observations at 01:54:16.2\,UT. At 6 minutes after the burst trigger, the Nordic Optical Telescope (NOT) equipped with the Andalucia Faint Object Spectrograph and Camera (AlFOSC) observed the GRB in the $R$-band. A sequence of $3\times600$\,sec spectra with a mean time of 2:26\,UT using a $1\farcs3$ wide slit and grism\,4 (with spectral coverage from 3800--9200\,\AA\ and resolution $R=\lambda/\Delta\lambda=277$) were obtained with AlFOSC. A strong 2175\,\AA\ bump was immediately obvious \citep{heintz18}. Another AlFOSC 2000\,sec spectrum was taken with grism\,20 (with spectral coverage from 6000--9200\,\AA\ and $R=560$) using a $1\farcs3$ slit.

The Gamma-Ray burst Optical and Near-Infrared Detector (GROND) mounted at the 2.2-m Max Planck Gesellschaft (MPG) telescope started taking photometric observations of the afterglow in all seven ($g^\prime r^\prime i^\prime z^\prime JH$ and $K$) bands at $\sim$1.05\,hrs after the burst. The complete log of GROND, NOT and VLT/X-shooter photometric observations is provided in Table \ref{tab:phot}. The AB magnitudes provided in Table \ref{tab:phot} are not corrected for Galactic extinction. From the photometry we construct multiband optical/NIR lightcurves. The temporal behaviour of the $r^\prime$-band lightcurve (between 0.3--8.3\,hrs) follows a broken power-law with a break at $t_b=6.48\pm1.20$\,ks and decay slopes $\alpha_1=0.02\pm0.03$ and $\alpha_2=2.00\pm0.13$ following the convention where $F_\nu\propto\nu^{-\alpha}$. The lightcurves of all other bands follow a consistent behaviour.

\begin{figure}
	\centering 
	{\includegraphics[width=\columnwidth,clip=]{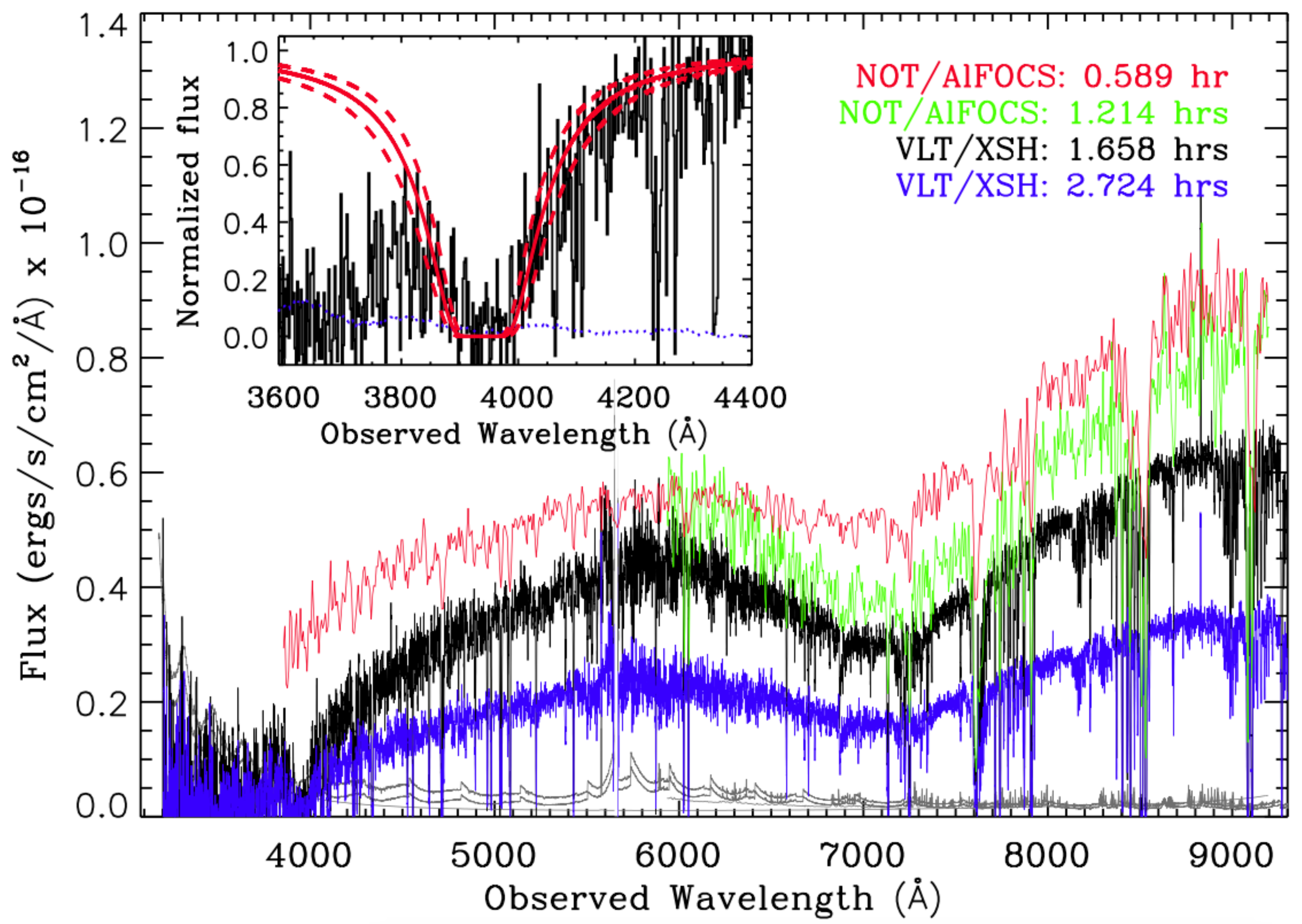} }
	\caption{The optical afterglow spectra of GRB\,180325A (in the extinction bump region) at four epochs observed with the NOT/AlFOSC and VLT/X-shooter instruments. Each color corresponds to a different epoch and mid-times of observations are indicated in the corresponding color. $inset$: Shown is the VLT/X-shooter spectrum and corresponding noise spectrum (blue dotted line) in the Ly$\alpha$ absorption-line region. The red solid and dashed lines show the best-fit {H}\,{\sc i} column density profile and 1$\sigma$ error, respectively.} 
		\label{obs} 
\end{figure}

We subsequently observed GRB\,180325A at $\sim$1.84\,hours after trigger with the VLT/X-shooter spectrograph. VLT/X-shooter data were obtained as part of the Stargate VLT proposal (Program ID: 0100.D$-$0649(A)) and the spectra were reduced following the approach outlined in \cite{selsing18}. The VLT/X-shooter spectrograph covers the spectral wavelength range from 3000--24,800\,\AA, divided into three spectroscopic arms. The spectroscopic sequence consisted of 2 epochs of $4\times 600$\,sec exposure each and were observed following an ABBA nodding pattern. The slits were aligned with the parallactic angle, and used slit widths are $1\farcs0$, $0\farcs9$ and $0\farcs9$ for the UVB, VIS and NIR arm, respectively. This corresponds to a spectral resolution of 5,400 (UVB), 8,900 (VIS), and 5,600 (NIR). The spectra were taken under good conditions with a median seeing of $1\farcs0$ at 6700\,\AA\ and at airmass ranging from 1.52--1.75. Throughout the paper, the wavelengths are reported in vacuum and are shifted to the heliocentric velocity. All optical data were corrected for a Galactic extinction of $E(B-V) = 0.02$\,mag using the dust maps of \cite{schlafly11}. As it is only a small correction, the uncertainty due to the Galactic extinction correction is negligible. All four, NOT and VLT/X-shooter, spectra in the extinction bump region are shown in Fig. \ref{obs}. The central redshift obtained from metal lines in the VLT/X-shooter spectrum is $z=2.2486$ and the derived {H}\,{\sc i} column density is log\,N({H\,{\sc i}}/cm$^{-2}$) $= 22.30\pm0.14$ (see Fig. \ref{obs} inset). The spectrum also shows the presence of a strong intervening system at $z = 2.04$.

\begin{table}
\caption{Multi-band photometric observations of GRB\,180325A.     
\label{tab:phot}}
\centering     
\setlength{\tabcolsep}{1pt}
\begin{tabular}{c c c c}  
\hline \hline                      
Facility-filter-band & Time since trigger & Exposure time & Observed magnitude \\
 & sec & sec & AB (mag) \\
\hline 
NOT-$R$ & 568.2 & 100.0 & $18.51\pm0.04$  \\
NOT-$R$ & 3091.2 & 100.0 & $19.23\pm0.04$ \\
GROND-$g^\prime$ & 3791.114 & 184.9 & $20.24\pm0.03$ \\
GROND-$r^\prime$ & 3791.114 & 184.9 & $19.32\pm0.03$ \\
GROND-$i^\prime$ & 3791.114 & 184.9 & $18.95\pm0.03$ \\
\hline
\end{tabular}
\tablecomments{Table \ref{tab:phot} is published in its entirety in the machine-readable format. A portion is shown here for guidance regarding its format and content.}
\end{table}

The X-ray spectra and lightcurves were extracted in the 0.3--10\,keV range following the method of \citet{evans09}. The entire XRT photon counting (PC)-mode dataset does not show strong spectral variations, thus justifying a fit with a single power-law. The best-fit is with a photon index $\Gamma=1.84\pm0.10$ and host galaxy equivalent neutral hydrogen column density, $N_{H,X}=1.33^{+0.42}_{-0.41}\times 10^{22}$\,cm$^{-2}$ where the Galactic {H}\,{\sc i} column density is set to $1.23\times 10^{20}$\,cm$^{-2}$ and Galactic molecular H$_2$ column density to $6.27\times 10^{18}$\,cm$^{-2}$ using the $nH$ tool\footnote{\url{http://www.swift.ac.uk/analysis/nhtot/}} following the method of \citet{willingale13}. We construct four XRT spectra around the SED epochs for the SED fitting (see \S\ref{results}). These four spectra are normalized to the photometric SED mid-points by considering the photon-weighted mean times and using the light curve decay power-law ($\alpha_X=2.10^{+0.07}_{-0.06}$) fit. The XRT data show no evidence of spectral evolution around the four spectra so the lightcurve hardness-ratio does not deviate from the mean.

\section{Results}\label{results}
\subsection{SED fitting and extinction curve}
We followed the SED fitting method described in \citet{zafar18}. We used the spectral fitting package \texttt{XSPEC} (v12.9; \citealt{arnaud96}) to fit the multi-epoch restframe NIR to X-ray SEDs of GRB\,180325A. The SEDs were modelled with a single or broken power-law plus \citet{fm90} parametric extinction law and soft X-ray absorption. We refer the reader to \citet{zafar18} for a more detailed description of the model. Briefly, the \citet{fm90} dust law consists of: $i)$ a UV linear component given as $c_1$ (intercept) and $c_2$ (slope) with $c_4$ defining the far-UV curvature and $ii)$ a Drude component specifying the 2175\,\AA\ extinction feature by $c_3$ (bump strength), $x_0$ (central wavelength), and $\gamma$ (bump width) parameters. The {\it total} Galactic equivalent neutral hydrogen column density ($N_{H,Gal}$) was fixed to the values obtained from \citet{willingale13}, $1.29\times 10^{20}$ cm$^{-2}$. The soft X-ray absorption, $N_{H,X}$, is left as a free parameter and modeled with {\tt ztbabs}. 

\begin{figure}
	\centering 
	{\includegraphics[width=\columnwidth,clip=]{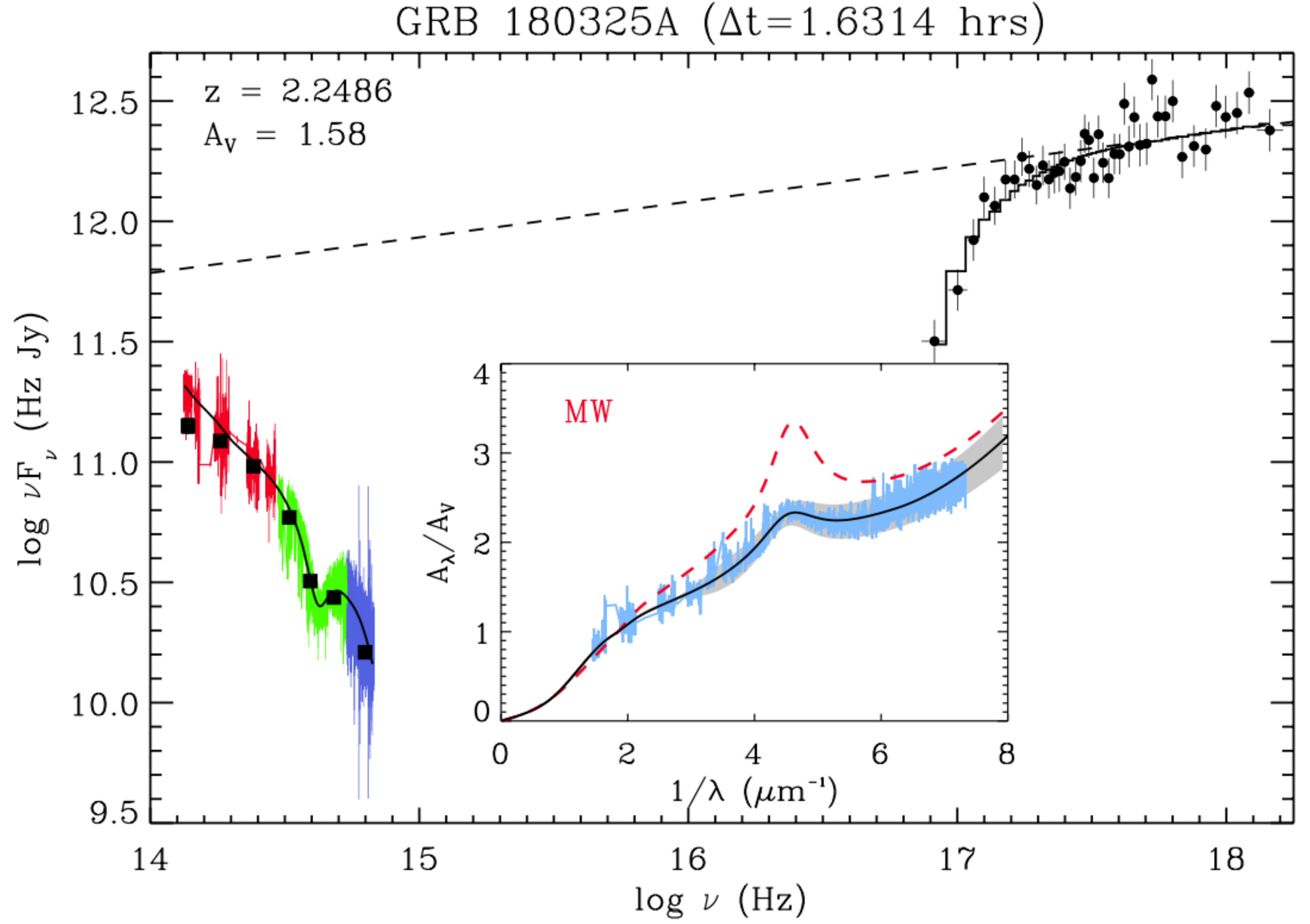} }
	\caption{Observer-frame afterglow SED of GRB\,180325A. Black circles indicate the \emph{Swift} X-ray data. The binned (for visual purposes) blue, green, and red data-points (with 1$\sigma$ errors) correspond to the UVB, VIS, and NIR VLT/X-shooter spectra, respectively. GROND observations are represented by black squares. The black solid and dashed lines show the best-fit dusty$+$absorbed and intrinsic spectral model, respectively. {\it Inset:} The black line shows the absolute extinction curve and the gray-shaded area materializes the associated 1$\sigma$ uncertainty. The X-shooter spectra are indicated by cyan curves. The red dashed line corresponds to the mean MW curve from \citet{fm07}.} 
		\label{sed180325A} 
\end{figure}

Except for the case of GRB\,080605 (two epochs; \citealt{zafar12}), the 2175\,\AA\ bump for GRB afterglows has not been spectroscopically observed at more than one epoch in any case. We here for the first time spectroscopically observed the extinction bump at four epochs. We used GROND photometric data and lightcurves (only for first epoch) to construct the X-ray to NIR SEDs around all four spectroscopic epochs. The X-ray data and optical spectra are scaled to the photometric mid-points. For X-shooter SEDs, the GROND data is only used for the normalization of spectra where difference is usually $<10\%$. For NOT SEDs, the GROND data outside the spectral coverage is incorporated in the SED fitting. We find that the SEDs at all epochs are fit well with a single power-law and a 2175\,\AA\ extinction curve (see, however, notes in Table \ref{best-fit}). We also tested if bump is related to the intervening absorber at $z=2.04$ and find no evidence. The SED and extinction curve at $\Delta$t $=1.6314$\,hrs are shown in Fig. \ref{sed180325A}. The best-fit results of the SEDs are reported in Table \ref{best-fit}. We find that for all four epochs, the extinction curve parameters are consistent, particularly the total-to-selective extinction ($R_V$) and the total extinction at $V$-band ($A_V$) are consistent within 1$\sigma$.

\begin{longrotatetable}
\begin{deluxetable*}{l c c c c c c c c c c c c c}
\tablecaption{Results of the SED best-fit parameters for GRB\,180325A X-ray to NIR SEDs.\label{best-fit}} 
\tablewidth{100pt}
\tabletypesize{\scriptsize}
\startdata
SED & $\Delta$t & $N_{\rm H,X}$ & $\beta$ & $c_1$ & $c_2$ & $c_3$ & $c_4$ & $\gamma$ & $x_0$ & $R_V$ & $A_V$ & $\chi^2_\nu$/dof & NHP\% \\
& hrs 	& 10$^{22}$ cm$^{-2}$ &   &  $\mu$m 	& & & $\mu$m$^2$ & $\mu$m$^{-1}$ & $\mu$m$^{-1}$ &  & mag & \\ 
\hline
NOT$-$XRT\footnote{Note that the spectrum and early photometry are affected by the host contribution. The XRT and multiband photometric data are also not available at that time and SED is generated from lightcurves extrapolation. The lightcurves show signs of evolution, therefore, results should not be considered reliable.} & 0.5833 & 0.59$^{+0.31}_{-0.29}$ & $0.76\pm0.13$ & $-1.92\pm0.21$ & $1.38\pm0.16$ & $3.10\pm0.21$ & $1.20\pm0.25$ & $1.14\pm0.03$ & $4.540\pm0.02$ & $4.40^{+0.23}_{-0.19}$ & $1.63^{+0.12}_{-0.10}$ & 1.00/881 & 51 \\
NOT$-$XRT & 1.1953 & 0.61$^{+0.30}_{-0.29}$ & $0.78\pm0.12$ & $-2.17\pm0.27$ & $1.31\pm0.23$ & $3.34\pm0.25$ & $1.21\pm0.47$ & $1.15\pm0.03$ & $4.540\pm0.02$ & $4.38^{+0.21}_{-0.18}$ & $1.55^{+0.11}_{-0.09}$ & 0.96/887 & 78 \\
XSH$-$XRT & 1.6314 & 0.70$^{+0.22}_{-0.19}$ & 0.85$^{+0.14}_{-0.13}$  & $-1.95\pm0.39$ & $1.28\pm0.17$ & $2.92\pm0.19$ & $0.52\pm0.14$ & $1.16\pm0.06$ & $4.538\pm0.03$ & $4.58^{+0.37}_{-0.39}$ & $1.58^{+0.10}_{-0.12}$ & 0.96/33572 & 100 \\
XSH$-$XRT & 2.9866 & 1.20$^{+0.37}_{-0.30}$ & 0.89$^{+0.12}_{-0.10}$ & $-2.04\pm0.34$ & $1.26\pm0.24$ & $3.05\pm0.22$ & $0.32\pm0.21$ & $1.13\pm0.05$ & $4.534\pm0.03$ & $4.46^{+0.41}_{-0.37}$ & $1.37^{+0.14}_{-0.10}$ & 0.99/33291 & 90 \\
\enddata
\end{deluxetable*}
\tablecomments{The columns give the SED photometric mid-point, the equivalent neutral hydrogen column density, intrinsic slope $\beta$, UV linear intercept $c_1$, UV slope $c_2$, bump strength $c_3$, far-UV curvature $c_4$, bump width $\gamma$, bump central wavelength $x_0$, $R_V$, visual extinction, reduced $\chi^2$ with number of degrees of freedom (dof) and null hypothesis probability (NHP).}
\end{longrotatetable}

\subsection{Metal absorption lines}
\begin{figure}
	\centering 
	{\includegraphics[width=\columnwidth,clip=]{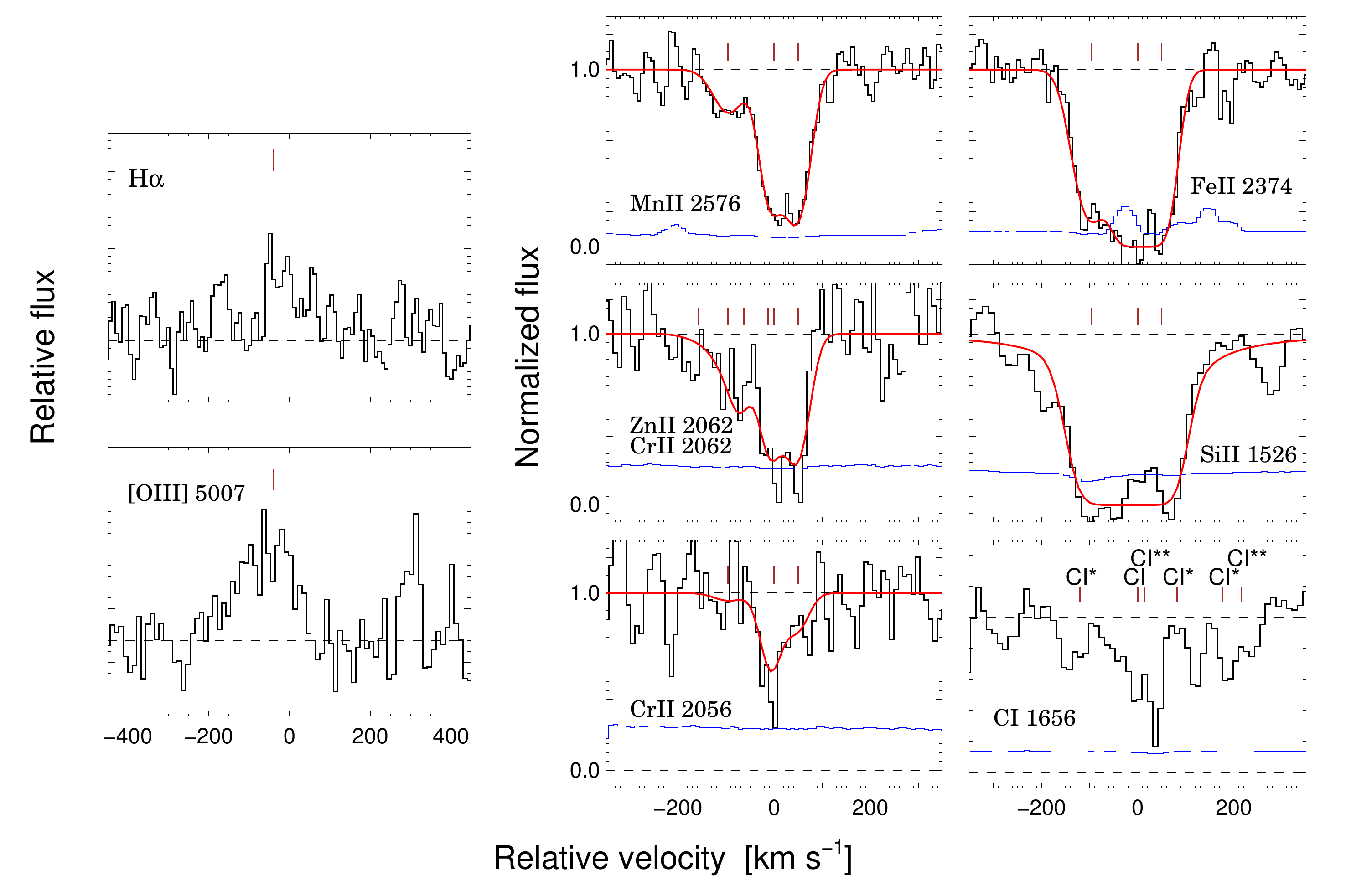} }
	\caption{{\it Left column:} Host galaxy H$\alpha$ and [O\,\textsc{iii}] emission lines. The central redshift of the GRB ($z=2.2486$) is marked with vertical brown lines. {\it Middle and right column:} Normalized absorption line profiles on a velocity scale centered at $z=2.2486$. The spectra are shown as black curves with corresponding error spectra in blue. Voigt-profile fits are shown in red and top vertical brown lines represent different velocity components.}
		\label{180325Avp} 
\end{figure}

Through a Ly$\alpha$ voigt profile fit we derive a column density of log\,N({H\,{\sc i}}/cm$^{-2}$) $= 22.30\pm0.14$ for atomic neutral hydrogen along the GRB sightline. To measure the gas-phase metallicity of the host galaxy, we fit multi-component Voigt profiles to the low-ionization metal lines in the spectrum using \textsc{VoigtFit} \citep{krogager18}. We show the best-fit Voigt profiles in Fig. \ref{180325Avp} to some of these metal lines. The derived column densities and abundances are reported in Table~\ref{tab:met} where solar photospheric abundances from \cite{asplund09} are assumed. The metal lines are all heavily saturated (except for Cr\,\textsc{ii} and C\,\textsc{i}), therefore, we report those abundances as lower limits. Note that Cr is depleted onto dust, and therefore its abundance is not indicator of the true metallicity. Using the column density of Zn, we derive a lower limit of $\rm [Zn/H] \geq-0.98$ for the metallicity of the host galaxy. We also use the low-ionization lines and measure an upper limit of $\Delta v_{90} \leq 180$\,km\,s$^{-1}$ for the velocity width of neutral gas in the host galaxy.


\begin{table}
\caption{GRB\,180325A metallicity with respect to Solar. The system has a three-component profile.}      
\label{tab:met} 
\centering     
\setlength{\tabcolsep}{1pt}
\begin{tabular}{c c c c c c}  
\hline \hline                      
Elemental &  Component 1 & Component 2 & Component 3 & Total & Metallicity \\
transition & log $N_M $ (cm$^{-2}$) & log $N_M$ (cm$^{-2}$) & log $N_M$ (cm$^{-2}$) & log $N_M$ (cm$^{-2}$) &[M/H] \\
 & $b=37.53\pm2.9$ km\,s$^{-1}$ & $b=20.9\pm5.5$ km s$^{-1}$ & $b=24.5\pm2.7$ & $\cdots$ & \\
\hline 
{H}\,{\sc i} & $\cdots$ & $\cdots$ & $\cdots$ & $22.30\pm0.14$ & $\cdots$ \\
{Fe}\,{\sc ii} & $14.87\pm0.07$ & $16.65\pm0.18$: & $15.31\pm 0.23$ & $16.68\pm 0.18$: & $> -1.35$: \\
{Mn}\,{\sc ii}& $12.94\pm0.06$ & $13.56\pm0.08$ & $13.73\pm 0.08$ & $13.99\pm 0.08$ & $> -1.90$ \\
{Zn}\,{\sc ii}& $13.13\pm0.14$ & $13.66\pm0.17$ & $13.81\pm0.15$ & $14.09\pm 0.15$ & $> -0.98$ \\
{Si}\,{\sc ii}& $14.97\pm0.37$ & $15.18\pm0.28$ & $16.03\pm 0.52$: & $16.12\pm0.48$: & $> -2.19$: \\
{Cr}\,{\sc ii}& $12.79\pm1.13$ & $13.75\pm0.18$ & $13.39\pm 0.37$ & $13.94\pm 0.30$ & $-2.00\pm0.33$ \\
\hline
\end{tabular}
\end{table}

\subsection{Neutral Carbon}
Observationally the presence of a 2175\,\AA\ extinction feature is found to be correlated with the detection of neutral atomic carbon \citep{zafar12,ledoux15,ma18} and/or molecules. This is consistent with a link between carbonaceous grain growth and the extinction bump \citep{henning98}. For a sample of quasar absorbers, \citet{ledoux15} found that $\sim$30\% of the {C}\,{\sc i} absorbers show a 2175\,\AA\ bump. Recently \citet{ma18} found that all quasar absorbers with {C}\,{\sc i} column densities above $\sim$10$^{14}$\,cm$^{-2}$ show a detectable bump feature in their SEDs. Based on a small sample, \citet{zafar12} reported a trend between {C}\,{\sc i} equivalent widths and bump strengths. In the afterglow spectrum of GRB\,180325A, we detect strong {C}\,{\sc i} absorption lines with rest-frame equivalent widths of $W_{\mathrm{r}}(\lambda 1560)=0.58\pm0.05$ and $W_{\mathrm{r}}(\lambda 1656) =0.85\pm0.05$. Comparing $W_{\mathrm{r}}$({C}\,{\sc i}) and the area of the bump ($A_{\rm bump}=\pi c_3/2\gamma$; at the 1.6314\,hrs epoch it is $3.95\pm0.20$), the present burst follows the trend presented in \citet{ma18}. 

\subsection{Host galaxy properties}
In the VLT/X-shooter NIR spectrum, we clearly detect H$\alpha$ and [O\,\textsc{iii}] emission lines from the GRB host galaxy. We measure line fluxes of $F_{{\rm H}\alpha}=(5.0\pm1.2)\times10^{-17}$erg\,s$^{-2}$\,cm$^{-1}$\AA$^{-1}$ and $F_{[{\rm O}\,\textsc{iii}]}=(7.7\pm0.9)\times10^{-17}$erg\,s$^{-2}$\,cm$^{-1}$\AA$^{-1}$ for these bright emission lines. The H$\alpha$ line flux corresponds to a star-formation rate (SFR) of $15.8\pm3.8$\,M$_\odot$yr$^{-1}$, based on the  calibration from \citet{kennicutt98}, and without correcting for dust extinction. Correcting for SED-derived dust extinction ($A_V\sim1.5$), the host galaxy SFR is $46\pm4$\,M$_\odot$yr$^{-1}$. We measure a Full-Width-Half-Maximum (FWHM) of $106\pm28$\,km\,s$^{-1}$ for the H$\alpha$ line. Using the correlation between stellar mass and velocity width of the H$\alpha$ line derived for GRB host galaxies in \citet{arabsalmani18}, we estimate a stellar mass of $M_\ast/M_\odot\sim 10^{9.3 \pm 0.4}$ for the host galaxy of GRB\,180325A. The estimated stellar mass and SFR place the host galaxy amongst the main-sequence star-forming galaxies with similar redshifts, though towards higher SFRs on the envelope of the main-sequence SFR-$M_\ast$ relation \citep[see Fig. 1 of][]{rodighiero11}. A larger specific SFRs (SFR per unit mass) than average is shown to be typical of GRB host galaxies \citep[see][and references therein]{arabsalmani18}.

%
%
\section{Discussion\label{discussion}}
We compared $R_V$, $A_V$, area of the bump ($A_{\rm bump}$), and maximum height ($E_{\rm bump}=c_3/\gamma^2$) of GRB\,180325A with other unambiguously detected GRB cases (GRB\,070802 and GRB\,080607: \citealt{zafar11}, GRB\,080605 and GRB\,080805: \citealt{zafar12}). Based on the 1.6314\,hrs SED results, we find this burst has wider bump area ($A_{\rm bump}=3.95\pm0.20$), height ($E_{\rm bump}=2.17\pm0.20$), and $R_V$ than all known GRB cases but has a lower dust content when compared to GRB\,080607 ($A_V=2.33^{+0.46}_{-0.43}$; \citealt{zafar11}). However, the GRB\,180325A bump area and height are still not as prominent as the mean MW \citep[$A^{{\rm MW}}_{\rm bump}=4.74$ and $E^{{\rm MW}}_{\rm bump}=3.04$;][]{fm07} and LMC average \citep[$A^{{\rm LMC}}_{\rm bump}=4.57\pm0.24$ and $E^{{\rm LMC}}_{\rm bump}=3.12\pm0.17$;][]{gordon03} bump. The extinction curve of GRB\,180325A appears to be shallower than the mean MW \citep{fm07} and average LMC \citep{gordon03} curves.

Diffuse interstellar bands (DIBs) are a ubiquitous phenomenon in Galactic sightlines, and have been detected in a limited number of extragalactic sightlines (Magellanic Clouds, supernovae, quasar absorbers). Remarkably, the one SMC sightline that exhibits normal strength DIBs includes a 2175\,\AA\ feature \citep{cox07}. The DIB carriers likely are large carbon-bearing molecules; C$_{60}^{+}$ has been proposed as the carrier of two (out of about 400) DIBs at 9577 and 9632\,\AA\ \citep{campbell15}. The measured (anomalous) extinction in GRB\,180325A, and the presence of the 2175\,\AA\ feature, motivates the search for the presence of the strongest DIBs in its VLT/X-shooter spectrum. Several of them happen to be redshifted into the opaque regions in between the NIR windows, therefore, we cannot find a significant detection.

The presence of CO and H$_2$ molecules is usually correlated with the UV bump \citep[e.g.,][]{prochaska09,ledoux15}. The spectrum blueward of Ly$\alpha$ is strongly suppressed and due to low signal-to-noise around the blue end of spectrum, there is no clear detection of H$_2$ molecules. Neutral chlorine could also be used as a tracer of H$_2$ \citep{balashev15}. The afterglow spectrum of GRB\,180325A has relatively low signal-to-noise in the {Cl}\,{\sc i} ($\lambda$1347\,\AA) region. The derived 3$\sigma$ limit of log\,N({Cl\,{\sc i}}/cm$^{-2}$) $<14.20$ suggests log\,N(H$_2$/cm$^{-2}$) $<20.60$ based of correlation presented in \citet{balashev15}. We also find no convincing detection of CO molecular lines because of low signal-to-noise and partly the spectral regions are blended with the strong intervening system at $z=2.04$.

\section{Conclusions\label{conclusions}}
We here report the detection of  a 2175\,\AA\ extinction bump in the spectrum of GRB\,180325A afterglow at $z=2.2486$, the only unambiguous detection over the past decade. For the first time, the extinction bump was spectroscopically observed at four different epochs twice with the NOT/AlFOSC and twice with the VLT/X-shooter. We constructed the X-ray lightcurve from \emph{Swift}-XRT and optical/NIR photometric lightcurves from GROND, NOT, and VLT/X-shooter. Using these multiband data, we generated SEDs at all four spectroscopic epochs. The SEDs fit well with a single power-law and a 2175\,\AA\ extinction bump with $R_V\approx4.4$, $A_V\approx1.5$. The bump and extinction curve of GRB\,180325A are shallower than the mean MW and LMC law. The derived metallicity of GRB is [Zn/H]$>-0.98$. The VLT/X-shooter spectrum exhibits neutral carbon at the GRB redshift, suggesting strong {C}\,{\sc i} is required to yield a 2175\,\AA\ bump. The dust-corrected SFR from the H$\alpha$ emission-line flux is $\sim46\pm4$\,M$_\odot$yr$^{-1}$ and the predicted stellar mass is log\,M$_\ast$/M$_\odot\sim9.3\pm0.4$ which is representative for long-duration GRB hosts.

\acknowledgements
Based on observations made with the Nordic Optical Telescope, operated by the Nordic Optical Telescope Scientific Association at the Observatorio del Roque de los Muchachos, La Palma, Spain, of the Instituto de Astrofisica de Canarias. Based on observations collected at the European Organisation for Astronomical Research in the Southern Hemisphere under ESO programme 0100.D$-$0649(A), PI: Tanvir. This work made use of data supplied by the UK Swift Science Data Centre at the University of Leicester. We acknowledge the use of public data from the \emph{Swift} data archive. DAK acknowledges support from the Spanish research project AYA 2014-58381-P, and from Juan de la Cierva Incorporaci\'on fellowship IJCI-2015-26153. RLCS acknowledges support from STFC. KEH and PJ acknowledge support by a Project Grant (162948--051) from The Icelandic Research Fund. The Cosmic Dawn Center is funded by the DNRF. LC is supported by DFF--4090-00079. SC, PDA and BS acknowledge support from the ASI grant I/004/11/3. AdUP acknowledges support from a Ram\'on y Cajal fellowship RyC-2012-09975 and the Spanish research project AYA 2014-58381-P

\bibliography{bump.bib}

\end{document}